\documentclass[twocolumn,prl,floatfix,superscriptaddress]{revtex4}
\usepackage{verbatim}
\usepackage{amsmath}
\usepackage{amssymb}
\usepackage{graphicx}

\usepackage{multirow}\usepackage{array}\usepackage{float}

\begin{document}

\title{Tunneling Spectroscopy of Quasiparticle Bound States in a Spinful Josephson Junction}

\author{W.~Chang}

\affiliation{Department of Physics, Harvard University, Cambridge, Massachusetts
02138, USA}

\affiliation{Center for Quantum Devices, Niels Bohr Institute, University of Copenhagen,
Universitetsparken 5, 2100 Copenhagen \O, Denmark}

\author{V.~E.~Manucharyan}

\affiliation{Society of Fellows, Harvard University, Cambridge, Massachusetts
02138, USA}

\author{T.~S.~Jespersen}

\affiliation{Center for Quantum Devices, Niels Bohr Institute, University of Copenhagen,
Universitetsparken 5, 2100 Copenhagen \O, Denmark}

\author{J.~Nyg{\aa}rd}

\affiliation{Center for Quantum Devices, Niels Bohr Institute, University of Copenhagen,
Universitetsparken 5, 2100 Copenhagen \O, Denmark}

\author{C.~M.~Marcus}

\affiliation{Department of Physics, Harvard University, Cambridge, Massachusetts
02138, USA}

\affiliation{Center for Quantum Devices, Niels Bohr Institute, University of Copenhagen,
Universitetsparken 5, 2100 Copenhagen \O, Denmark}

\begin{abstract}
The spectrum of a segment of InAs nanowire, confined between two superconducting leads, was measured as function of gate voltage and superconducting phase difference using a third normal-metal tunnel probe. Sub-gap resonances for odd electron occupancy---interpreted as bound states involving a confined electron and a quasiparticle from the superconducting leads, reminiscent of Yu-Shiba-Rusinov states---evolve into Kondo-related resonances at higher magnetic fields. An additional zero-bias peak of unknown origin is observed to coexist with the quasiparticle bound states.
\end{abstract}
\maketitle

A quantum spin impurity in a superconductor modifies the order parameter locally via exchange interaction with quasiparticles, thereby creating Yu-Shiba-Rusinov sub-gap states~\cite{yu1965,shiba1968,rusinov1969,soda1967,shiba1969para, balatsky2006impurity}.
For weak exchange interaction, a sub-gap state near the gap edge emerges from singlet correlations between the impurity and the quasiparticles. Increasing exchange interaction lowers the energy of the singlet state and increases a key physical parameter, the normal state Kondo temperature $T_K$. At $k_BT_K\sim\Delta$ (Kondo regime), where $\Delta$ is the superconducting gap, the energy gain from the singlet formation can exceed $\Delta$, resulting in a level-crossing quantum phase transition (QPT)~\cite{balatsky2006impurity,satori1992numerical,simonin1995rare,bauer2007spectral}. The QPT changes the spin and the fermion parity of the superconductor-impurity ground state, and is marked by a peak in tunneling conductance at zero bias~\cite{franke2011competition}. 

Yu-Shiba-Rusinov states can be studied in a controlled manner with a mesoscopic quantum-dot/superconductor Josephson junction [Fig.~1(a) and 1(b)]~\cite{de2010hybrid}. Such a device provides a novel control knob that tunes the exchange interaction---the superconducting phase difference across the junction, $\phi$. We label the states of the device $\left|n_{dot},n_{lead}\right\rangle$, where $n_{dot}$ and $n_{lead}$ represent the electron/quasiparticle occupation numbers in the dot/superconducting leads. When the spin orientations of the states are important, we replace the occupancy numbers with arrows. First, a spin 1/2 impurity is created by trapping a single electron in the lowest available orbital of the dot (assuming large level spacing) with a Coulomb barrier [Fig.~1(c)]~\cite{goldhaber1998kondo,Buitelaar2002Quantum}. This barrier is maximal at the electron-hole (e-h) symmetry point, where the spinful state, $|1,0\rangle$, costs less than both the empty, $|0,0\rangle$, and the doubly occupied, $|2,0\rangle$, states by the charging energy $U$. 
Charge fluctuation at energies below $\Delta$ is suppressed for $U>\Delta$.
Next, spin-flip scattering connects the degenerate states $\left|\uparrow,\downarrow\right\rangle $ and $\left|\downarrow,\uparrow\right\rangle$ via the virtual population of states $|2,0\rangle$ [Fig.~1(d)] or $|0,0\rangle$ [Fig.~1(e)]. These two scattering channels cause an effective (super-) exchange interaction between quasiparticles and the spinful dot. Finally, 
the exchange interaction is sensitive to phase due to the quantum interference of the spin-flip scattering amplitudes between the two leads.
Compared to scattering via $|2,0\rangle$, scattering via $|0,0\rangle$ differs by a phase factor $\mathrm{exp}(-\imath\phi)$ because it is accompanied by a Cooper pair transfer [Fig.~1(e)].
At $\phi=\pi$ these two scattering channels interfere destructively, making the exchange coupling minimal at $\phi=\pi$ and maximal at $\phi=0$. Consequently, both the singlet excited state, $|S\rangle$, and the doublet ground state, $|D\rangle$, acquire a phase modulation, albeit only in higher order processes for the latter~\cite{glazman1989resonant,spivak1991negative,rozhkov2001josephson,martin2011josephson,Rozhkov2000Interacting,Vecino2003Josephson,tanaka2007kondo,Meng2009}.

\begin{figure}[b!]
\center \label{figure1}
\includegraphics[width=3in]{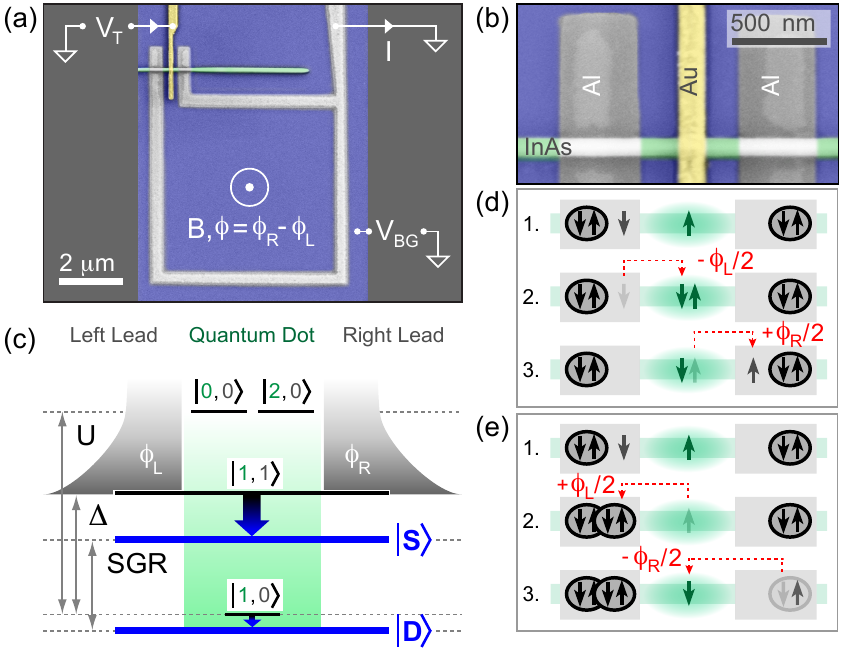}
\caption{\footnotesize{(color online). (a), (b) Scanning electron micrographs of a lithographically identical device. (c) Lowest energy states of a single-orbital quantum dot at the electron-hole symmetry point for $k_{B}T_{K}\ll\Delta$. The states are labeled by their electron/quasiparticle occupation number in the format $|n_{dot}, n_{lead}\rangle$. Exchange interaction dresses the states $|1,0\rangle$ and $|1,1\rangle$ as the doublet, $|D\rangle$, and the singlet, $|S\rangle$, states respectively. Transition from $\left|D\right\rangle $ to $\left|S\right\rangle$ produces a sub-gap resonance (SGR). (d), (e) Phase sensitive spin-flip processes coupling the $|1,1\rangle$ states $\left|\uparrow,\downarrow\right\rangle$ and $\left|\downarrow,\uparrow\right\rangle$ via virtual occupation of (d) $\left|2,0\right\rangle $ and (e) $\left|0,0\right\rangle $. 
}}
\end{figure}

Previous experiments have investigated the ground state of spinful Josephson junctions~\cite{van2006supercurrent,cleuziou2006carbon,jorgensen2007critical,eichler2009tuning,maurand2012first,Luitz2012Understanding}. Phase-biased junctions with weak coupling showed negative supercurrent~\cite{van2006supercurrent,cleuziou2006carbon}, consistent with theoretical predictions of the weak phase-modulation of $|D\rangle$~\cite{glazman1989resonant,spivak1991negative,rozhkov2001josephson}, while for strong coupling, positive supercurrent was observed~\cite{eichler2009tuning,jorgensen2007critical}. The latter was interpreted in terms of a QPT associated with the interchange of states $\left|S\right\rangle $ and $\left|D\right\rangle$ at $k_{B}T_{K}\sim\Delta$~\cite{eichler2009tuning,maurand2012first,Luitz2012Understanding}. Meanwhile, other experiments have performed tunneling spectroscopy on spinful Josephson junctions without phase-control~\cite{nadyagraphene,deacon2010tunneling,deacon2010ZBP, lee2012ZBP}, or with phase-control but away from the Kondo regime~\cite{pillet2010andreev}. This leaves the effect of phase on sub-gap states in the Kondo regime unaddressed. Tunneling spectroscopy in similar devices has also been used recently to examine signatures of Majorana end states~\cite{mourik2012majorana,das2012majorana,deng2012majorana}. 

In this Letter, we demonstrate both phase and gate control of sub-gap states in a Kondo-correlated Josephson junction ($k_BT_K\sim\Delta$)~\cite{de2010hybrid}. We also report the first evidence of a singlet/doublet QPT induced by the superconducting phase difference. Our junction is an InAs nanowire contacted by a pair of superconducting Al leads and a third normal metal tunnel probe. At magnetic fields above the critical field of Al, tunneling into the InAs quantum dot with odd electron occupancy showed Kondo resonances~\cite{goldhaber1998kondo} with associated Kondo temperatures, $T_{K}\sim 1$~K. Near zero field, tunneling into the nanowire revealed the superconducting gap of the Al leads, $\Delta\simeq 150\,\mu\mathrm{eV}$, and a pair of sub-gap resonances (SGR) symmetric about zero bias. Both gating the nanowire and varying the phase bias across the junction tune the SGRs. For certain parameters, the two SGRs cross at zero bias, which we interpret as a level-crossing QPT. However, no such crossing occured upon suppressing $\Delta$ to zero with an applied magnetic field. Instead, the SGRs evolve smoothly into Kondo resonances, and this transition is typically accompanied by the appearance of a separate zero-bias resonance of unknown origin.

\begin{figure}[t!]
\label{figure2}
\includegraphics[width=3.2 in]{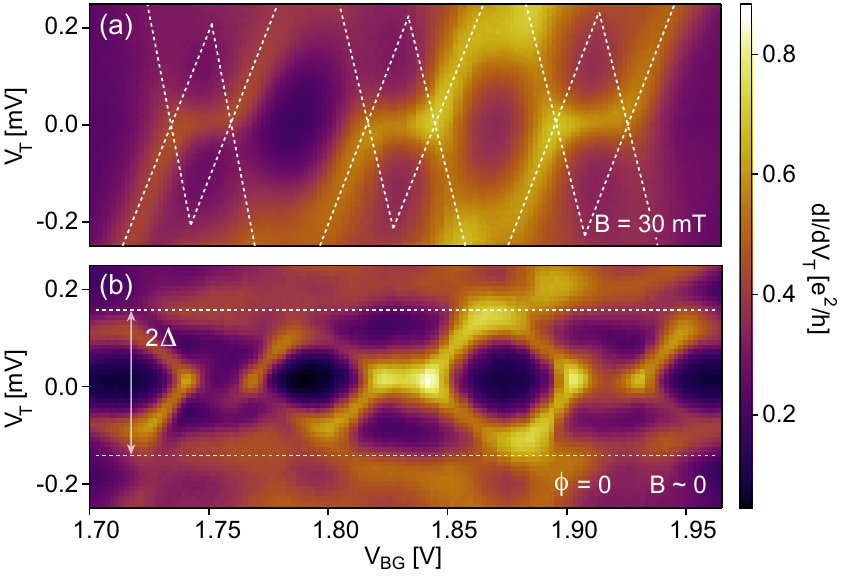}
\caption{\footnotesize{(color online). Differential conductance as a function of tunnel-probe voltage, $V_\mathrm{T}$, and back-gate voltage, $V_{\mathrm{BG}}$. (a) Normal state data, $B$~=~30~mT. (b) Superconducting state data, $B$~$\sim$~0 and $\phi$~=~0. Coulomb diamonds in (a) and superconducting gap in (b) are highlighted with dotted lines.}}
\end{figure}

Our device consisted of a 10~$\mu$m long, 100~nm diameter InAs nanowire contacted by two ends of a superconducting loop (5/100~nm Ti/Al). A third normal metal tunnel probe (5/100~nm Ti/Au) contacted the nanowire between the two superconducting leads [Figs.~1(a) and 1(b)]. The segment of nanowire between the superconducting contacts was 0.5~$\mu$m long, with a loop area of $\sim25\,\mu\mathrm{m}^{2}$. For this loop area, the flux period, $h/2e$, corresponds to a perpendicular magnetic field period of $72\,\mu\mathrm{T}$. The nanowires were grown by molecular beam epitaxy on (111)B-oriented InAs substrates using Au nanoparticles as catalyst, resulting in a wurtzite structure with a [111] growth direction. They were then liberated from the InAs growth substrate by sonication in methanol, and deposited on a degenerately doped Si substrate with a 100~nm thermal oxide. Highly transparent barriers between Al and InAs were made using an ammonium polysulfide etch immediately before deposition~\cite{suyatin2007sulfur}. Lower transparency of the normal contact was achieved by etching for less time. The device was measured in a dilution refrigerator with a base temperature 20 mK, through several stages of low-pass filtering and thermalization. 

When superconductivity in the entire device was suppressed by an applied magnetic field $B$, diamond patterns characteristic of weak Coulomb blockade (CB) were observed in transport between the loop and the normal lead [Fig.~2(a)]. Consecutive diamonds alternate in size, indicating that the orbital level spacing, $\xi$, is comparable to the charging energy, $U\simeq200\,\mu\mathrm{eV}$. The smaller (odd occupancy) diamonds contain gate-independent zero-bias ridges that split at higher magnetic fields (Sup.~1), typical of the Kondo effect~\cite{goldhaber1998kondo,jespersen2006kondo}. From the temperature dependence of the zero-bias ridges, we estimate $T_{K}$ to be in the range of 0.5-1~K (Sup.~3). Poor visibility of the odd diamonds suggests strong coupling to the superconducting leads ($\Gamma_{S}\gtrsim U$), and the amplitudes of the Kondo ridges indicate an asymmetry between superconducting and normal contacts~\cite{kretinin2011spin}. The estimated asymmetry, $\Gamma_{N}\lesssim\Gamma_{S}/10$, allows qualitative treatment of the Au lead as a weak tunnel probe. 

In the superconducting state ($B$~$\sim$~0), gap-related features were observed at tunnel-probe voltages, $V_\mathrm{T}\simeq\pm150\,\mu V\simeq \pm\Delta/e$, consistent with the gap of Al. SGRs symmetric about zero bias were also observed [Fig.~2(b)]. Comparison of Figs.~2(a) and 2(b) shows that the positioning (in backgate voltage, $V_\mathrm{BG}$) of SGRs in the superconducting state coincide with CB and Kondo features in the normal state. In particular, the SGRs and their symmetric partners converge towards each other, to the extent that sometimes they overlap, in an odd CB valley. In contrast, they are pushed towards the gap edge in the even CB valleys. 

\begin{figure}[t!]
\center\label{figure3} 
\includegraphics[width=3.2 in]{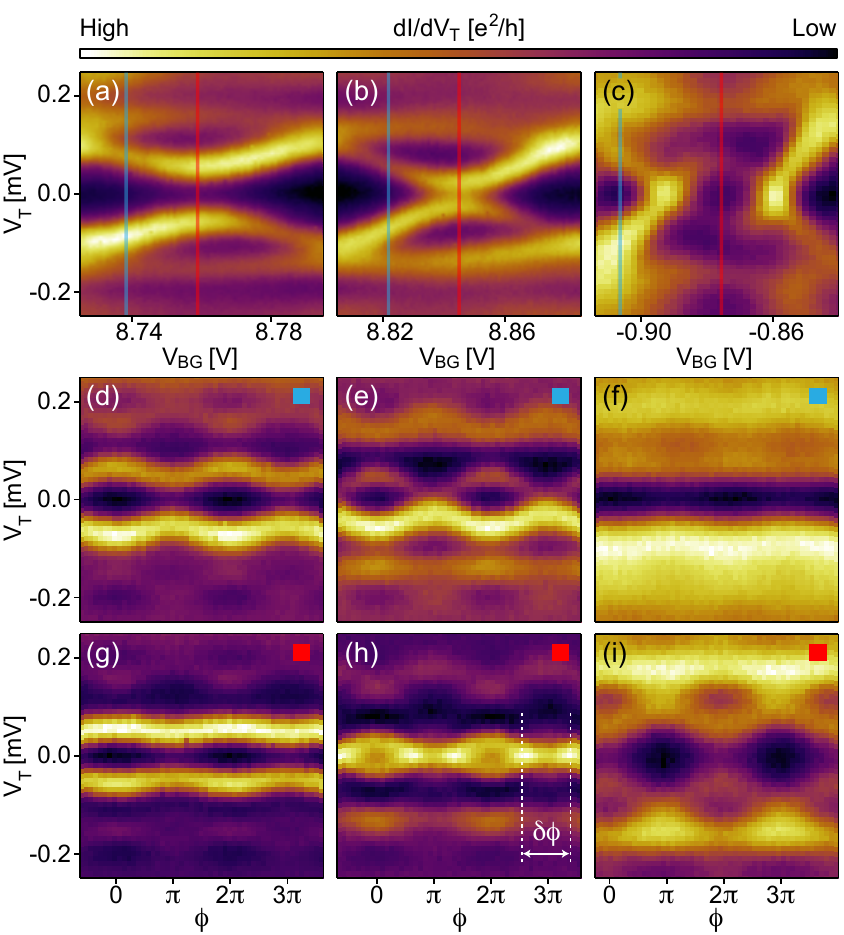}
\caption{\footnotesize{(color online). Three sub-gap resonances (SGRs) arranged in columns of increasing  Coulomb energy. (a)--(c) Back-gate dependence of the SGRs at $\phi=0$. The lower rows show their corresponding phase dependence off (d)--(f) and on (g)--(i) the electron-hole symmetry point. (d)--(g) Conventional phase dependence. (h) Hybrid phase dependence. (i) $\pi$-shifted phase dependence}}
\end{figure}

Based on their qualitative dependence on $V_\mathrm{BG}$ and $\phi$, three categories of SGRs in the case of a spinful dot were identified: (i) For small charging energy, $U<(\Delta,\Gamma_{S}$), SGRs do not cross the zero-bias axis for any $V_{\mathrm{BG}}$ or $\phi$ [Figs.~3(a), 3(d), and 3(g)]. The SGR energy is maximal at $\phi=0$ and minimal at $\phi=\pi$ [Fig.~3(d) and 3(g)]---this is the conventional phase dependence of non-interacting Josephson junctions~\cite{WChangUnpublished}. (ii) For large charging energy, $U >\Delta$, (Sup.~2) SGRs become inverted, crossing zero bias twice as a function of $V_{\mathrm{BG}}$ [Fig.~3(c)]. Between zero-bias crossings, the phase dependence of SGR energies is the opposite of the conventional behavior, that is, minimal at $\phi=0$ and maximal at $\phi=\pi$ [Fig.~3(i)]. We call this a $\pi$-shifted phase dependence. Outside the intersections in $V_{\mathrm{BG}}$, the phase dependence of SGR energy is conventional [Fig.~3(f)]. (iii) For moderate charging energy $U\sim\Delta$ [Figs.~3(b), 3(e), and 3(h)], SGRs do not intersect for any $V_{\mathrm{BG}}$ at $\phi = 0$ [Fig.~3(b)]. Phase dependence away from the e-h symmetry point is conventional [Fig.~3(e)], but close to the symmetry point, the pair of SGRs intersect twice per phase period of $2\pi$ [Fig.~3(h)]. Crossings occur at $\phi=\pi\pm\delta\phi/2$, where $\delta\phi<\pi$ is the phase-difference between the two closest crossings [Fig.~3(h)]. With this type of SGR, the phase dependence depends on the phase value itself: it is conventional for $\phi\sim0$, and $\pi$-shifted for $\phi\sim\pi$. 

\begin{figure*}[t!]
\center\label{figure4} 
\includegraphics[width=6.5 in]{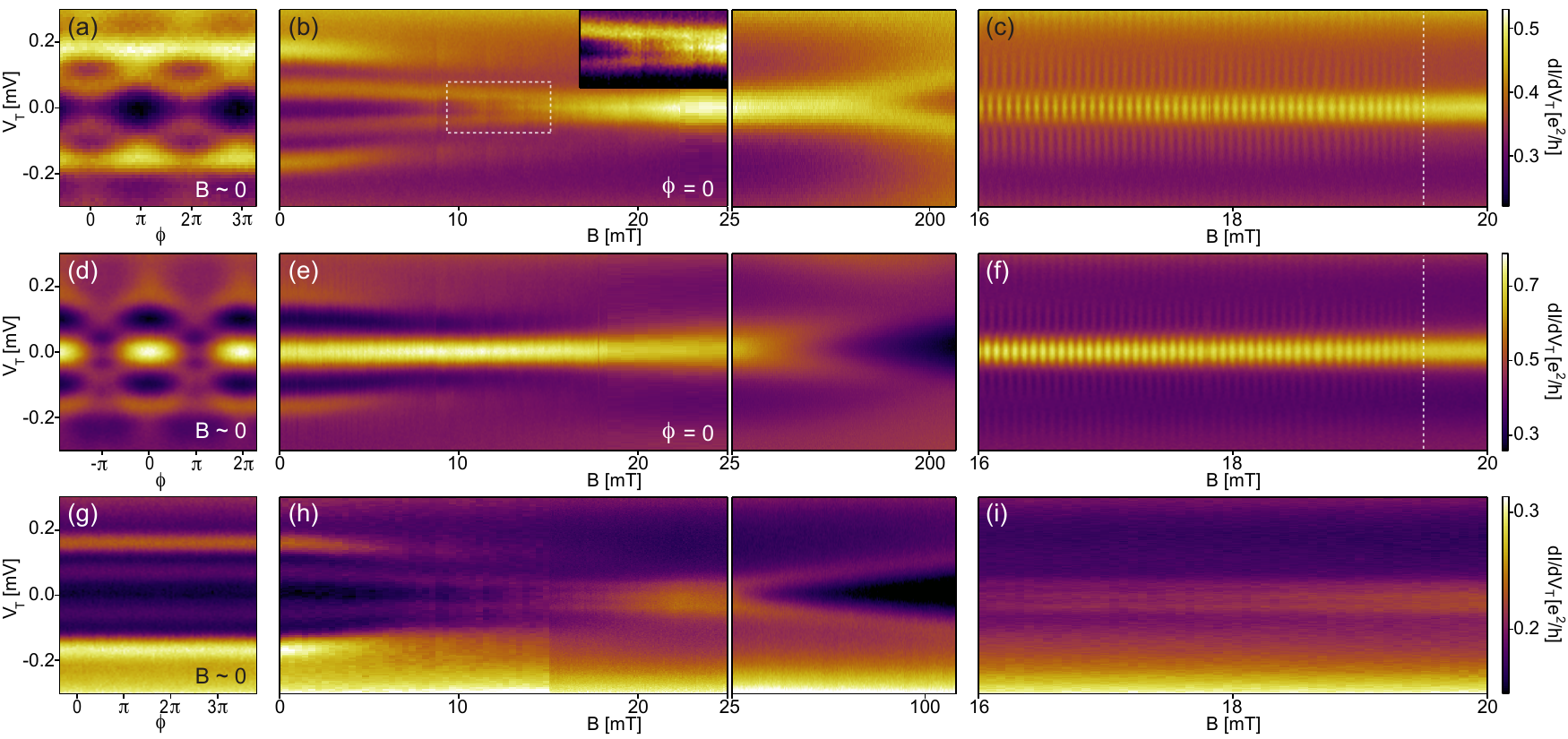}
\caption{\footnotesize{(color online). Arranged in the order of decreasing $T_{K}$, each row shows the evolution of a SGR at the electron-hole symmetry point as a function of phase and magnetic field. The left column shows phase dependence at $B$~$\sim$~0, the center column shows magnetic field dependence at $\phi=0$, and the right column shows the magnetic field and phase dependence around $B$~=~18~mT. To obtain the phase constant panels (b) and (e), we select $\phi=0$ data points from the full data set. The oscillations of the SGRs disappear abruptly at $B$ = 19.5 mT (dotted lines) in both (c) and (f). Inset in (b) is a closeup of the region outlined with dotted lines. A third resonance, pinned at zero bias, is clearly visible in the high contrast color scale.}}
\end{figure*}

Focusing on $\pi$-shifted SGRs, we examine in Fig.~4 the magnetic field evolution of three SGRs measured at their e-h symmetry points. The first SGR [Figs.~4(a)--4(c)] is the same as in Fig.~3(c). Selecting $\phi=0$ from the full data set (Sup.~9), the well separated and resolved SGRs gradually approach zero-bias and merge into a Kondo resonance in the normal state [Fig.~4(b)]. Temperature dependence of the normal-state Kondo peak gives $T_K\simeq1\,\mathrm{K}$~\cite{goldhaber1998kondo} (Sup.~3). Taking $g \sim 13$ from normal-state CB data (Sup.~4), the splitting of the Kondo peak at $\sim 140$ mT is consistent with this value of $T_K$~\cite{quay2007magnetic} [Fig.~4(b)]. In the other two cases (middle and lower rows of Fig.~4), Kondo peaks split at lower fields of $B\sim50$~mT [Fig.~4(e)] and $B<20$~mT [Fig.~4(h)], suggesting lower Kondo temperatures.

In the second case [Figs.~4(d)--4(f)], SGRs overlap at zero-bias for $\phi=0$, but are separated for $\phi=\pi$ [Fig.~4(d)]. The overlapping SGRs at zero field evolve continuously into a Kondo resonance as the field is increased into the normal-state regime [Fig.~4(e)]. Phase dependent oscillations of the SGR vanish abruptly at a critical value of field, $B_{c}=19.5\,\mathrm{mT}$ [Fig.~4(f)]. 
The same critical field is observed in Fig.~4(c), 
and also in higher density regimes of the device (Sup.~5). 

The last case in Fig.~4 has no phase-dependence [Fig.~4(g)], presumably because of poor coupling to one of the superconducting contacts. However, its $V_{BG}$ dependence allows us to establish that this SGR is indeed a $\pi$-shifted type (Sup.~8). Here, in contrast to the first two cases, the pair of SGRs evolve continuously and directly into a split Kondo peak without ever merging or crossing at zero bias [Figs.~4(h) and 4(i)].

Close inspection of Fig.~4 reveals an unexpected and intriguing feature: a narrow needle-like resonance pinned at zero bias. In Fig.~4(b), this ``needle'' is absent at $B=0$ but appears for $B>10$~mT while the leads are still superconducting. In Fig.~4(d) the needle is hidden by the SGRs at $\phi=0$, yet it is clearly visible at $\phi=\pi$. In this case, the needle  exists at $B=0$, and merges into the normal-state Kondo resonance at higher field [Sup.~9(f)]. In Fig.~4(h), the needle appears at $B>10$~mT, similar to the case in Fig.~4(b), despite a large difference in Kondo temperatures. In fact, the strength of the needle appears uncorrelated with $T_K$ of the normal-state Kondo peak (Sup.~10). The needle is also distinct from the normal state Kondo resonance as seen in Figs.~4(h) and 4(i), where three separate peaks can be identified: The two peaks flanking the central needle appear to emerge from the SGR at the low-field end and evolve continuously into the split Kondo peaks at the the high-field end. We find that the needle only appears between the two $V_\mathrm{BG}$ intersection points of $\pi$-shifted SGRs, which in turn corresponds to an odd Coulomb diamond (Sup.~6). Finally, the needle appears brighter at $\phi=0$, when the separation between the two SGR is the smallest [Figs.~4(c) and 4(d)] (Sup.~6).

We now compare theoretical expectations for SGRs~\cite{martin2011josephson} to experimental observations. At the e-h symmetry point of a spinful quantum dot with suppressed charge fluctuations, the phase-tunable exchange interaction detaches a singlet state $|S\rangle$ down from the gap edge [Fig.~1(c)]. Since quantum interference weakens the exchange interaction at $\phi=\pi$ [Figs.~1(d) and 1(e)], a $\pi$-shifted SGR is indeed expected (phase modulation of the energy of $|D\rangle$, being a higher-order effect, is much weaker than that of $|S\rangle$)~\cite{Rozhkov2000Interacting,Vecino2003Josephson,tanaka2007kondo,Meng2009}. This is consistent with out experiment, as seen, for example, in Fig.~3(i). Strong coupling to the leads, reflected in the large $T_{K}$, should further result in a SGR that is well separated from the gap edge at $\phi = 0$~\cite{simonin1995rare,bauer2007spectral}. Detuning $V_{\mathrm{BG}}$ towards a neighboring even diamond increases charge fluctuations and mixes either $\left|0,0\right\rangle $ or $\left|2,0\right\rangle $ into $|S\rangle$, thereby lowering its energy. Consequently, one expects a level-crossing QPT to a singlet ground state as $V_{\mathrm{BG}}$ approaches an even diamond, in agreement with the zero-bias crossings in Fig.~2(b) and Fig.~3(c). This QPT is predominantly governed by the enhanced charge flutuations away from the e-h symmetry spot. Finally, the observed conventional phase dependence in the even state of the dot [Fig.~3(f)] is also expected, because a spinless dot acts purely as a potential scatterer, behaving as an effectively non-interacting junction~\cite{Koerting2010Nonequilibrium}.

A more interesting QPT occurs in Fig.~3(h) as a function of phase-bias. It corresponds to a situation, where the energy gain from the quasiparticle-dot singlet formation makes this state the ground state at $\phi=0$, but not at $\phi=\pi$. This behavior is known in the theory literature as $0^{'}$-junction or $\pi^{'}$-junction~\cite{rozhkov1999,martin2011josephson}, and, to our knowledge, has not been reported in previous experiments. 

Reducing $\Delta$ sufficiently below $k_BT_K$ should result in a level-crossing QPT that is driven entirely by spin fluctuations~\cite{balatsky2006impurity}. Experimentally, we would see a zero-bias crossing of the SGRs at $B<B_c$ as $B$ is increased to suppress $\Delta$. However, this theoretical expectation is not seen in our device as exemplified in Figs.~4(b) and 4(h), perhaps obscured by our current experimental resolution or by the needle feature. The needle may be related to similar features observed in recent experiments \cite{deacon2010ZBP,lee2012ZBP}. An unlikely soft gap in Al may explain such a resonance in terms of conventional Kondo screening. We note, however, that the needle itself does not split with increasing $B$, as one might expect from a conventional Kondo effect. More intriguingly, the needle appears much stronger at $\phi=0$ than at $\phi=\pi$, suggesting possible phase dependence and a link to the sub-gap states (Sup.~11). While the observed behaviors of sub-gap states agree at $B\sim0$ with existing theory on Yu-Shiba-Rusinov states, further theory and experiment are needed to understand the origin of the needle and the magnetic field dependence of the sub-gap states~\cite{domanski2008,oguri2012}.

We thank P.~Krogstrup, M.~Madsen, and C.~S\o rensen for MBE growth. We also thank D.~Abanin, J.~Bauer, H.~Churchill, E.~Demler, K.~Flensberg, L.~Glazman, K.~Grove-Rasmussen, F.~Kuemmeth, J.~Paaske, D.~Pekker, and J.~Sau for fruitful conversations. Support from the Carlsberg Foundation, the Danish National Research Foundation, the Lundbeck Foundation, and Microsoft Project Q is gratefully acknowledged.

\end{document}